\documentclass[3p,times]{elsarticle}
\usepackage{ecrc}
\usepackage[bookmarks=false]{hyperref}
    \hypersetup{colorlinks,
      linkcolor=blue,
      citecolor=blue,
      urlcolor=blue}

\volume{00}

\firstpage{1}

\journalname{Procedia Computer Science}

\runauth{A. Roccon}

\jid{procs}

\usepackage{listings}
\usepackage{amssymb}
\usepackage{amsthm}

\usepackage[figuresright]{rotating}

\begin{document}
\begin{frontmatter}



\dochead{Proceedings of the First EuroHPC user day}

\title{A GPU-ready pseudo-spectral method for direct numerical simulations of multiphase turbulence}


\author[a,b]{Alessio Roccon\corref{cor1}}

\address[a]{Polytechnic Department of Engineering and Architecture, University of Udine, 33100 Udine, Italy}
\address[b]{Institute of Fluid Mechanics and Heat Transfer, TU-Wien, 1060 Vienna, Austria}

\begin{abstract}
In this work, we detail the GPU-porting of an in-house pseudo-spectral solver tailored towards large-scale simulations of interface-resolved simulation of drop- and bubble-laden turbulent flows.
The code relies on direct numerical simulation of the Navier-Stokes equations, used to describe the flow field, coupled with a phase-field method, used to describe the shape, deformation, and topological changes of the interface of the drops or bubbles.
The governing equations -- Navier-Stokes and Cahn-Hilliard equations -- are solved using a pseudo-spectral method that relies on transforming the variables in the wavenumber space.
The code targets large-scale simulations of drop- and bubble-laden turbulent flows and relies on a multilevel  parallelism.
The first level of parallelism relies on the message-passing interface (MPI) and is used on multi-core architectures in CPU-based infrastructures.
A second level of parallelism relies on OpenACC directives and cuFFT libraries and is used to accelerate the code execution when GPU-based infrastructures are targeted.
The resulting multiphase flow solver can be efficiently executed in heterogeneous computing infrastructures and exhibits a remarkable speed-up when GPUs are employed.
Thanks to the modular structure of the code and the use of a directive-based strategy to offload code execution on GPUs, only minor code modifications are required when targeting different computing architectures.
This improves code maintenance, version control and the implementation of additional modules or governing equations.
\end{abstract}

\cortext[cor1]{Corresponding author. Tel.: +39-0432-558006.}
\end{frontmatter}

\email{alessio.roccon@uniud.it}


\section{Introduction}
\label{main}

Turbulent multiphase flows are ubiquitous in nature and our everyday life. 
These flows play a key-role in various applications, from geophysical phenomena \cite{Jahne1998,pereira2018reduced}, to environmental and industrial processes \cite{paul2004handbook,schramm2003,shaw2003particle}. 
With respect to single-phase turbulence, the description and modeling of multiphase turbulence is much more complex, since these type of flows require the description of an ever-moving and deforming interface, its topological modifications, and the underlying turbulent flow \cite{Elghobashi2018,cmmf2009,Roccon2023,soligo2021}.
Simulations constitute an essential tool to investigate the physics of multiphase turbulence and are becoming increasingly popular in recent years: numerical simulations give access to detailed space- and time-resolved information on the flow field and on the morphology of the two phases.

Obtaining an accurate description of the dynamics of a turbulent multiphase flow on a discretized temporal and spatial grid is however a challenging task because of the large scale separation that characterizes these flows: scales range from the largest flow scale (of the order of the domain size), down to the Kolmogorov scale of turbulence and further down to the molecular scale of the interface.
This has direct implications on the description of multiphase turbulence as the spatio-temporal resolution one can reasonably afford is limited by computing capabilities \cite{Gorokhovski2008}. 
Specifically, as done for single-phase turbulence \cite{kim1987,orlandi2000fluid,rogallo1984numerical}, it would be beneficial to perform simulations in which all scales are directly resolved, without any model.
However, this approach cannot be applied to multiphase flows, since the scale separation between the largest flow scale and the smallest interfacial scale is about eight to nine orders of magnitude, while the most recent high-performance computing (HPC) infrastructures, can handle a maximum scale separation of about three to four orders of magnitude. 

In the last 15 years, the use of graphics processor units (GPU) has played a key role in the development of modern HPC systems \cite{nickolls2010gpu,owens2008gpu} paving the way for big advances in many different fields \cite{dematte2010gpu,friedrichs2009accelerating,pandey2022transformational,pratx2011gpu}.
In computational fluid dynamics, a large amount of computing resources are usually required for complex flow problems, especially when direct numerical simulations of multiphase turbulence are involved.
Thanks to the integration of GPUs in HPC centers, there is now the possibility to solve large-scale problems \cite{ames2020multi,bernardini2023streams,costa2021gpu,de2023uranos,pirozzoli2021one,zhu2018afid}.
However, with respect to code development for CPUs, programming for GPUs poses new challenges: i) CPUs and GPUs are characterized by different computing architectures; ii) Specific programming languages/models and compilers are required to generate code that can be executed on GPUs; iii) Performance, scaling and code portability across different vendors and infrastructures represent an open issue; iv) CPU and GPU memories are physically separated and efficient strategy for the memory management should be designed \cite{fujii2013data}.
All these aspects render the development and optimization of algorithms on GPU-accelerated infrastructures an open challenge.
 
In this context, different approaches are available and they are characterized by different degrees of portability, adaptability, and performance. 
The most commonly used are: i) Compute Unified Device Architecture (CUDA) \cite{cuda}, an extension of Fortran and C/++ programming languages to offload computations on GPUs and to manage memory transfers; ii) Open Computing Language (OpenCL) \cite{opencl2010}, an open, royalty-free standard for cross-platform, parallel programming of diverse accelerators; iii) OpenMP \cite{dagum1998openmp} and OpenACC \cite{openacc2011openacc}, directives-based models that use directives to define data movement and computation offloading; iv) Library-based solution (e.g. Kokkos \cite{Kokkos2014}), which offers high-level abstract programming concepts for performance portability on different types of high-performance computing infrastructures; v) Standard language parallelism, a parallel programming model that allows for accelerating C++ and Fortran codes without using language extensions or additional libraries.
Each of these approaches is characterized by its own advantages and disadvantages from the point of view of the performance, portability, maintenance, short- and long-term programming language compatibility and maintenance, compiler requirements,  and future developments that one should carefully evaluate before starting the GPU-porting phase of a code.
With the idea of having a single code that can be executed on different architectures, we choose here to follow the directive-based approach offered by OpenACC.
Among the available approaches, this approach has the following advantages: i) A single code has to be maintained: OpenACC directives will be ignored when GPU support is not enabled; ii) In the framework of the Nvidia HPC Software Development Kit (SDK) it allows for the use of the managed memory feature, which greatly simplifies data transfer between CPU and GPU memories; iii) Additional features, like the solution of new governing equations, can be easily implemented as the development time required to port to GPUs new code sections is reduced (compared to the other available approaches).

In this work, the OpenACC programming model is applied to a code for direct numerical simulation of multiphase turbulence based on a pseudo-spectral method \cite{HussainiZ1987}.
In particular, the code relies on a direct solution of the Navier-Stokes equations coupled with a phase-field method to describe the interface shape and its topological changes.
The parallelization is based on a multilevel approach (MPI + X).
A first level relies on MPI and FFTW libraries \cite{frigo2005} and it is employed when the code is executed in CPU-based computing infrastructures. 
A second level of parallelism that employs OpenACC directives and cuFFT libraries \cite{cuFFT} is used to accelerate the code execution when GPU-based infrastructures are targeted.
Thanks to this combination of OpenACC directives and cuFFT libraries, all the most computationally intensive sections of the code can be offloaded to GPUs. 
Overall, when GPUs are used, this leads to a remarkable speed-up and thus a reduction of the wall-clock time required for a time step (with respect to when CPUs are used).

The paper is organized as follows. In Section 2, the governing equations are presented; in Section 3 the numerical method is detailed.
Then, in Section~4, the GPU-porting and the code performance are detailed and a demo simulation is presented.
Finally, conclusions are drawn in Section~5.

\section{Methodology}

\subsection{Governing equations}

To describe the dynamics of the system, direct numerical simulation (DNS) of the Navier-Stokes equations, used to describe the flow field, are coupled with a phase-field method (PFM), used to describe interfacial phenomena.
These equations are solved in a plane channel geometry: the streamwise and spanwise directions are periodic while along the wall-normal directions no-slip or free-slip, or a combination of them, can be applied.

\subsubsection{Phase-field method}

The phase-field method (PFM) is an interface-capturing method used for the description of multiphase flows.
The method is based on the introduction of a marker function -- the phase-field variable  $\phi$ -- that is uniform in the bulk of the phases  ($\phi=\pm1$) while it varies smoothly over the thin transition layer that separates the two phases \cite{anderson1998,jacqmin1999,Roccon2023}. 
The time evolution of the phase-field variable is here described by the Cahn-Hilliard (CH) equation which, in dimensionless form, reads as:
\begin{equation}
    \frac{\partial \phi}{\partial t}+\mathbf{u} \cdot \nabla \phi = \frac{1}{Pe} \nabla^2 \mu\, ,
    \label{eq:CHE}
\end{equation}
where $\mathbf{u}=( u, v, w)$ is the velocity vector, $\mu$ is the chemical potential.
The P{\'e}clet number, $Pe$, represents the ratio between the diffusive timescale, $h^2/\mathcal{M} \beta^2$, and the convective time scale, $h/u_{\tau}$:
\begin{equation}
    Pe=\frac{u_\tau h}{\mathcal{M} \beta}\, ,
\end{equation}
where $u_\tau=\sqrt{\tau_w/\rho}$ is the friction velocity (with $\tau_w$ the wall shear-stress and $\rho$ the density), $h$ is the half-channel height, $\mathcal{M}$ is the mobility parameter, and $\beta$ is a positive constant introduced in the dimensionless procedure. 

The chemical potential, $\mu$, is obtained as the variational derivative of the Ginzburg-Landau free-energy functional \cite{badalassi2003,jacqmin1999,Soligo2019c,toth2015}. 
To model the case of two immiscible fluids, the free-energy functional,  $\mathcal{F[\phi,\nabla \phi]}$, is composed by the sum of two different contributions \cite{badalassi2003,jacqmin1999}:
\begin{equation}
\mathcal{F[\phi,\nabla \phi]} = \int_\Omega \big( f_0 + f_{mix} \big) d \Omega\,  = \int_\Omega  \underbrace{ \frac{1}{4}(\phi^2-1)^2}_{f_0} +  \underbrace{\frac{Ch^2}{2} |\nabla \phi|^2 }_{f_{mix}} ,
\end{equation}
The first term, $f_0$ (bulk energy), identifies the tendency for the multiphase system to separate into two pure stable phases, while the second contribution,  $f_{mix}$ (mixing energy), is a non-local term accounting for the energy stored at the interface (i.e. surface tension).
The Cahn number, $Ch=\epsilon/h$, represents the dimensionless thickness of the interfacial layer separating the two phases.
By taking the functional derivative of the Ginzburg-Landau free-energy functional, we obtain the expression of the chemical potential:
\begin{equation}
    \mu = 
    \frac{\delta \mathcal{F[\phi,\nabla \phi]}}{\delta \phi}=\phi^3-\phi-Ch^2\nabla^2\phi\, .
\end{equation}
At equilibrium, considering that the chemical potential is constant throughout the domain (${\bf \nabla} \mu = {\bf 0}$), the following (equilibrium) profile is obtained for a planar interface: 
\begin{equation}
    \phi_{eq}=\tanh \left( \frac{s}{\sqrt{2} Ch} \right)\, ,
    \label{width}
\end{equation}
where $s$ is the coordinate normal to the interface (located at $s=0$).

\subsubsection{Hydrodynamics}

To describe the flow-field of the multiphase system, the Cahn-Hilliard equation~(\ref{eq:CHE}) is coupled with the Navier-Stokes equations \cite{Elghobashi2018,cmmf2009,soligo2021}.
We consider here two incompressible and Newtonian phases and, for the sake of simplicity, we assume that they are characterized by equal density, $\rho$, and viscosity, $\mu$.
With these assumptions, the Navier-Stokes equations can be written as:
\begin{equation}
    \nabla \cdot \mathbf{u}=  0 \, ,
    \label{cont} 
\end{equation}
\begin{equation}
   \frac{\partial \mathbf{u}}{\partial t} + \mathbf{u} \cdot \nabla \mathbf{u}  =   -\nabla p 
    + \frac{1}{Re_{\tau}} \nabla^2 \mathbf{u}    + \mathbf{f}_\sigma\, ,
\label{eq:NSE}
\end{equation}
where $p$ is the pressure-field and $\mathbf{f}_\sigma$ represents the surface tension forces.
These forces are here computed using a continuum-surface stress approach \cite{gueyffier1999volume,lafaurie1994modelling} :
\begin{equation}
\label{eq:kort}
\mathbf{f}_\sigma = \frac{3}{\sqrt{8}}\frac{Ch}{We}\nabla \cdot [\bar \tau_c ]\, ,
\end{equation}
where  $\bar \tau_c=|\nabla \phi|^2 \mathbf{I}-\nabla \phi \otimes \nabla \phi$ is the Korteweg tensor used to model surface tension forces \cite{korteweg1901}.

The dimensionless numbers that appear in the  Navier-Stokes equation are the friction Reynolds number:
\begin{equation}
    Re_\tau=\frac{\rho u_\tau h}{\eta}\, ,
\label{eq:Re}
\end{equation}
which represents the ratio between the inertial and viscous forces, and the Weber number:
\begin{equation}
We=\frac{\rho u_\tau^2 h}{\sigma}\ ,
\label{eq:We}
\end{equation}
which represents the ratio between inertial and surface tension forces (being $\sigma$ the surface tension).

\subsection{ Numerical method}

The Navier-Stokes and continuity equations are not solved in primitive variables (velocity and pressure) but are recast in the so-called wall-normal velocity-vorticity formulation.
This avoids solving a Poisson equation for pressure.
Specifically, Navier-Stokes and continuity equations are replaced by a set of four equations \cite{kim1987,soligo2019a,speziale1987}: i) A second-order equation for the wall-normal component of the vorticity; ii) A fourth-order equation for the wall-normal component of the velocity vector; iii) Continuity equation; iv) Definition of wall-normal vorticity. 
To obtain the set of governing equations for the wall-normal velocity-vorticity formulation, we can first rewrite the Navier-Stokes equations as follows:
\begin{equation}
\frac{\partial {\bf u}}{\partial t}={\bf S}-\nabla p+\frac{1}{Re_\tau}\nabla^2 {\bf u} \, ,
\end{equation}
where the term $\bf S$ contains all the non-linear terms present in the Navier-Stokes equations (e.g. convective terms, surface tension forces, etc.).
By taking the curl of the Navier-Stokes equations, the pressure term vanishes thanks to the identity $\nabla \times \nabla p =0$ and a transport equation for the vorticity vector, $\mathbf \omega$, is obtained:
\begin{equation}
\label{vort}
\frac{\partial {\mathbf{\omega}}}{\partial t}=\nabla \times {\bf S} +\frac{1}{Re_\tau}\nabla^2 {\bf \omega} \, ,
\end{equation}
By taking again the curl of the vorticity transport equation, we obtain the following $4$-$th$ order equation for the velocity:
\begin{equation}
\label{4thw}
 \frac{\partial \nabla^2 \mathbf{u}}{\partial t} = \nabla^2 \mathbf{S} - \nabla(\nabla \cdot \mathbf{S})  + \frac{1}{Re_\tau} \nabla^4 \mathbf{u}\, ,
\end{equation}
We solve here for the wall-normal components of the vorticity $\omega_z$ and velocity $w$.
Hence, the following governing equations are solved:
\begin{equation}
\label{omegaz}
\frac{ \partial \omega_z}{\partial t}  =  \frac{\partial S_y}{\partial x}   - \frac{\partial S_x}{\partial y} + \frac{1}{Re_\tau} \nabla^2 \omega_z\, ,
\end{equation}
\begin{equation}
  \label{ww}
 \frac{ \partial (\nabla^2 w) }{\partial t}  =  \nabla^2 S_z - \frac{\partial}{\partial z}\left(\frac{\partial S_x}{\partial x} + \frac{\partial S_y}{\partial y} + \frac{\partial S_z}{\partial z}\right)
                      + \frac{1}{Re_\tau} \nabla^4 w\, ,
\end{equation}
 complemented by the continuity equation~(\ref{cont}) and the definition of wall-normal vorticity:
\begin{equation}
\label{omegadef}
 \omega_z  =  \frac{\partial v}{\partial x} - \frac{\partial u}{\partial y}\, .
\end{equation}

The Cahn-Hilliard equation, which is a fourth-order equation, is split into two second-order equations \cite{badalassi2003}. 
In particular, the CH equation is first rewritten in the following way:
\begin{equation}\
\label{phisplit}
\frac{\partial \phi}{\partial t} = S_\phi + \frac{s Ch^2}{Pe_\phi} \nabla^2\phi - \frac{Ch^2}{Pe_\phi} \nabla^4\phi\, ,
\end{equation}
where  $S_\phi$ represents the contribution of the non-linear terms.
The operator splitting $\nabla^2\phi = \nabla^2 \phi (sCh^2+1) - s Ch^2 \nabla^2\phi$ is then applied \cite{badalassi2003} where the positive coefficient $s$
has been chosen considering the temporal discretization.

The governing equations~(\ref{cont})-(\ref{omegaz})-(\ref{ww})-(\ref{omegadef})-(\ref{phisplit}) are solved using a pseudo-spectral method: the variables are transformed from the physical into the wavenumber space. 
Along the periodic directions ($x$ and $y$), all the quantities are expressed by Fourier series while they are represented by Chebyshev polynomials along the non-homogeneous wall-normal direction.
The corresponding $N_x \times N_y \times N_z$ collocation points are equally spaced along the $x$ and $y$ directions while they are stretched along the wall-normal direction where a finer grid resolution is obtained near the two walls.
Thanks to the adoption of a pseudo-spectral discretization, the equations are decoupled along the two periodic directions.
In this way, a series of 1D Helmholtz independent problems along the wall-normal directions are obtained.
All calculations are carried out in the wavenumber space except the non-linear terms, which are first computed in the physical space and then transformed back to spectral space (pseudo-spectral method).
This avoids the evaluation of (computationally expensive) convolutions  in the wavenumber space \cite{HussainiZ1987,peyret2002}.
The governing equations are discretized in time using an IMplicit-EXplicit (IMEX) scheme, in which the non-linear terms are integrated with an Adams-Bashfort scheme, while the linear terms are integrated by a Crank-Nicolson (Navier-Stokes) or by an implicit Euler (Cahn-Hilliard) scheme.

To advance the velocity and phase-field variables from time step $n$ to time step $n+1$, the following intermediate steps are performed (hat notation is used to identify the variable representation in the wavenumber space).
\begin{itemize}
\item[i)] The velocity field, $\mathbf{u}^n$ and the phase-field, $\phi^n$, are initialized and variables are transformed in the spectral space, $\hat{\mathbf{u}}^n$ and $\hat{\phi}^n$ (where hat notation is used in the following to identify the spectral representation of the variables).
\item[ii)] The non-linear convective and surface tension terms, i.e. the term $\bf S$, are computed. 
As these terms are non-linear, they are computed in the physical space rather than in the spectral space via convolution operation.
\item[iv)] Continuity and Navier-Stokes equations are solved (using the wall-normal velocity-vorticity formulation) to obtain the new velocity field, $\hat{\mathbf{u}}^{n+1}$.
This requires the solution of a system of Helmholtz problems along the wall-normal direction for each ($x,y$) location.
\item[vi)] The non-linear terms present in the Cahn-Hilliard equation are computed, i.e. the term $S_\phi$ is formed. Also in this case, the non-linear terms are computed in the physical space.
\item[v)] The Cahn-Hilliard equation is solved and the new value of the phase-field, $\hat{\phi}^{n+1}$, is obtained.
Similar to the solution of the Navier-Stokes equations, this requires the solution of two Helmholtz problems along the wall-normal direction for each ($x,y$) location.
\end{itemize}

\section{Implementation}

This numerical scheme has been implemented in a parallel Fortran 2003 MPI in-house proprietary code (FLOW36).
The code is written using Fortran 2003 and the main parallelization backbone relies on an MPI paradigm. 
On top of the MPI backbone, OpenACC directives and cuFFT libraries are used to accelerate the code execution on GPU-based computing infrastructures.
The code has been largely used in the past for the investigation of a wide range of complex turbulent flows: particle-laden flows \cite{marchioli2002mechanisms}, stratified turbulent flows \cite{Zonta2012},  interface-resolved simulations of drop- and bubble-laden flows \cite{Soligo2019,Mangani2022,Roccon2023}, drag reduced flows \cite{roccon2024}.
The code is developed using a modular approach: conditional compilation directives are used to enable (or disable) the different governing equations/physics available.

\subsection{First level of parallelization: MPI}

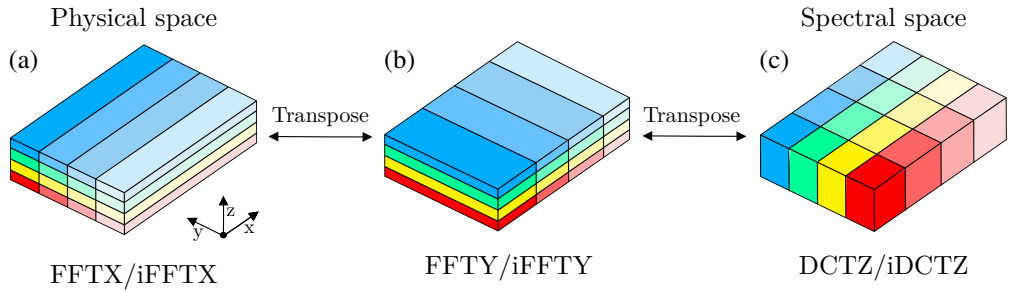
\begin{figure}
\center 
\setlength{\unitlength}{0.0025\columnwidth}
\begin{picture}(400,100)
\put(30,0){\includegraphics[width=0.8\columnwidth, keepaspectratio]{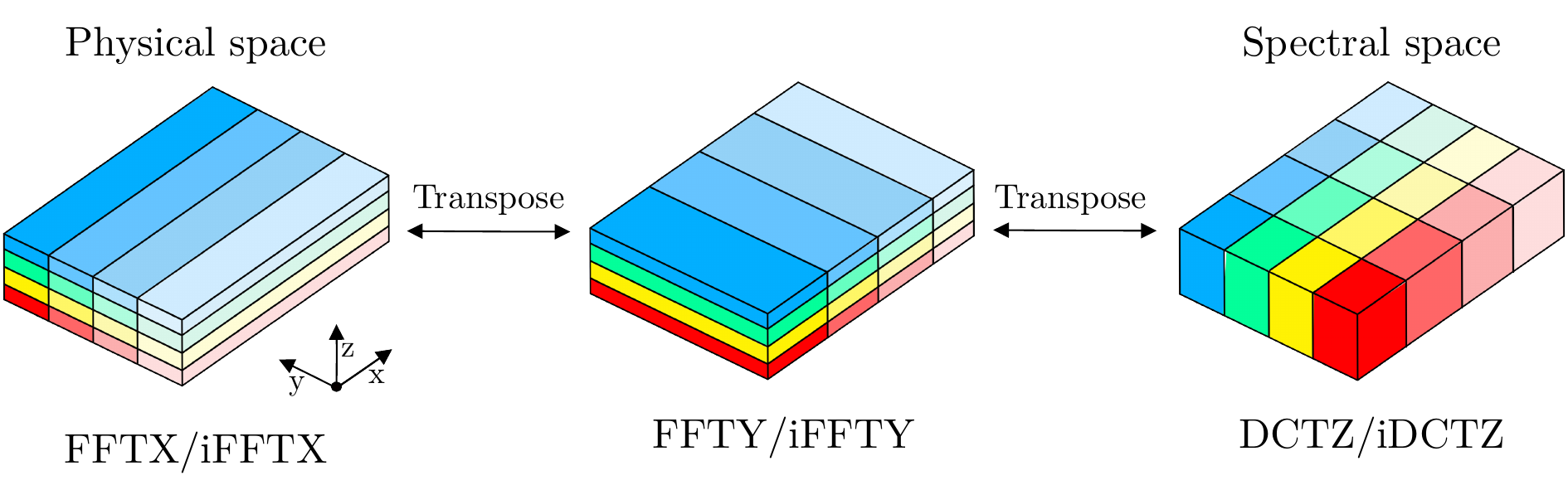}}
\put(30,75){(a)}
\put(150,75){(b)}
\put(270,75){(c)}
\end{picture}
\caption{2D domain decomposition employed for the first level of parallelization (MPI).
Each color corresponds to a different MPI task.
In physical space, the domain is divided along the $y$ and $z$ directions (pencils orientated along the $x$-direction), while in spectral space it is divided along the $x$ and $y$ directions (pencil orientated along the $z$-direction). 
Transpositions (i.e. loops of MPI communications) are required to change the pencil orientation and to compute the transform along the various directions: along $x$ in the first configuration shown in panel~$a$, along $y$ in the configuration shown in panel~$b$, and along $z$ in the configuration shown in panel~$c$.}
\label{mpi}
\end{figure}

The parallelization backbone of the code relies on an MPI approach; the overall workload is divided among the different MPI tasks using a 2D domain decomposition (pencil decomposition).
Within this strategy, the whole domain is split in so-called pencils: the domain is divided along two out of three directions and each sub-domain is assigned to a different MPI process.
In physical space, the domain is divided along the $y$ and $z$ directions (pencils oriented along the $x$-direction), while in modal space it is divided along the $x$ and $y$ directions (pencils oriented along the $z$-direction). 
This change in the pencil orientation is needed when taking the transforms: to compute the Fourier or Chebyshev transforms each process must hold all the points in the transform direction. 
Thus, when in physical space, at first Fourier transforms are taken in the $x$ direction (Figure~\ref{mpi}$a$), then the parallelization changes in order to have all the points along the $y$ direction. 
The domain is divided between the $x$ and $z$ directions when taking the Fourier transforms along $y$ (Figure~\ref{mpi}$b$).
At this point, each $x-y$ plane holds all the Fourier modes at a certain height. 
Then the parallelization is again changed, switching to a domain decomposition along $x$ and $y$ directions, thus each MPI process holds all the points in $z$ direction at a certain ($x,y$) location (parallelization in modal space). 
Finally, Chebyshev transforms are taken in the $z$ direction (Figure~\ref{mpi}$c$).

The only MPI communications occur when the non-linear terms of the governing equations are computed.
These terms are computed in the physical space and thus a change of pencil orientation is required to compute the transforms along the three directions (from spectral to physical and vice-versa).
After the calculation of all the non-linear terms, a system of Helmholtz problems along the wall-normal direction ($z$) is solved at each ($x,y$) location independently.

\subsection{Second level of parallelization: OpenACC directives}

On top of the MPI parallelization scheme presented above, cuFFT libraries and OpenACC directives are used to accelerate the code execution. 
When CPU-based architectures are employed, each MPI task is assigned to a physical core.
Differently, when GPU-based architectures are used, each MPI task is assigned to a specific GPU.
All the computationally intensive operations are performed on the GPUs. 
Specifically, the cuFFT libraries are used to perform all the transforms (Fourier and Chebyshev) and the entire solver can be efficiently executed on GPUs thanks to the fine-grain parallelism offered by the numerical scheme (series of 1D independent problems along the wall-normal direction). 
To obtain a code that can be easily ported between CPU- and GPU-based architectures, we employ the managed memory model present in OpenACC (which exploits the CUDA unified memory feature).
As in most systems CPU and GPU memories are physically separated, the use of GPUs usually requires explicit memory transfers and the handling of the copies shared between CPU and GPU memories.
Differently, using the managed memory feature, there is no need to explicitly define each memory transfer between the CPU (host) and the GPU (device).
In particular, memory can be accessed using a single pointer from CPU or GPU code. 
All memory transfers are managed by the compiler that makes the decisions about when and how to move data from the host to the device and back.
The use of the managed memory model should be specified via the \texttt{-gpu=managed} compiling option.
Although its use greatly simplifies code optimization, GPU-porting, and implementation of new features, one should bear in mind the three main drawbacks of this approach: i) The actual bandwidth of the memory transfer between CPU and GPU is slightly reduced; this is due to the virtual memory page faulting and migration used by the compiler to migrate data; ii) Data cannot be transferred asynchronously (note that asynchronous compute
regions can be created); iii)  This feature requires the use of Nvidia compiler and is at the moment non-compatible with GPUs from different vendors.


\subsubsection{Fourier and Chebyshev transforms:  cuFFT libraries}

To accelerate the execution of Fourier and Chebyshev transforms, these operations are performed using the highly optimized and scalable cuFFT libraries.
This CUDA-based library is optimized for Nvidia GPUs and offers an interface equivalent to that employed by the FFTW library (for CPUs) \cite{frigo2005design}.
In FLOW36, conditional compilation flags are used to compile the code with the correct library (FFTW when using CPUs and cuFFT when using GPUs).
To perform a transform, three main steps are required: i) Creation of a plan; this operation set-ups the library and identifies the best algorithm to be used depending on the available resources; ii) Execution of the transform according to the pre-defined plan; iii) Destruction of the plan. 
Plans can be also stored so that 
All transforms are executed using the batched version of the library (specified by the plan): this means that the library is set to perform a certain number of transforms along one direction and the input data is coherent with the advanced data layout structure.
For instance, considering the forward transform along the $x$ direction, $N_y \times N_z$ transforms of a signal having length $N_x$ are performed.
The transform can be then executed as follows by invoking the cuFFT library via OpenACC.
For instance, considering the forward transform along $x$, this operation can be performed as follows:
\begin{lstlisting}
!$acc data copyin(input) copyout(output)
!$acc host_data use_device(input,output)
gerr=gerr+cufftExecD2Z(plan,input,output)
!$acc end host_data
!$acc end data
\end{lstlisting}
where data clauses are required to force the data movement between CPU and GPU memories when libraries are used.
The suffix \texttt{D2Z} defines the type of transform: as here we consider the forward transform along $x$, the input is an array of double (\texttt{D}) while the output is an array of complex numbers (\texttt{Z}), i.e. a real-to-complex transform.
The above-mentioned strategy can be straightforwardly applied to the Fourier transform along $x$ and $y$, which correspond to the first and third index of a generic 3D input array.

A different approach is required to perform the Chebyshev transforms along the wall-normal direction $z$.
Two minor issues arise here: i) The cuFFT libraries do not explicitly support real-to-real transforms; ii) We need fo perform a transform along the direction specified by the inner index of a matrix and this does not allow for the use of the batched mode of cuFFT.
These two issues can be easily overcome by manipulating the input and output vector so that a Chebyshev transform is performed and by transposing (before and after the transform) the input vector so that it satisfies the advanced data layout.
In particular, the Chebyshev transform can be performed using FFT algorithms as it belongs to the group of discrete sine/cosine transforms (specifically, a DCT-I).
The Chebyshev transform of an N-point real signal can be obtained by taking the discrete Fourier transform of a 2N-point even extension of the signal \cite{makhoul1980fast}.
This scheme is applied to both the real and imaginary parts of the input signal and for both forward and backward transforms (the backward transform of a DCT-I transform is a DCT-I \cite{saverin2023sailffish}).
Note indeed that the input is a complex array because the FFTs along the $x$ and $y$ directions have been already performed at this stage, see figure~\ref{mpi}c.
Although the present strategy involves two transposes (back-and-forth) for each transform, the possibility to perform the Chebyshev transform in batched mode outperforms the other possible approach, i.e. a recursive loop of single 1D transforms, by one order of magnitude in terms of performance.
Further optimization of the Chebyshev transform step can be obtained using the strategy proposed by \citet{makhoul1980fast}, which does not require performing a transform of size $2N$ but only of size $N$ by re-ordering the input signal.

\subsubsection{Solver execution: OpenACC directives}

The computation of the non-linear terms in physical space, which mainly requires matrix multiplications, has been offloaded to GPUs by means of OpenACC directives. 
In particular, considering the computation of the non-linear terms present in both the Navier-Stokes and Cahn-Hilliard equations, this operation can be straightforwardly parallelized on GPUs as it mainly involves element-wise product (Hadamard product).
Indeed, all the flow-field and phase-field variables are collocated and thus defined on the same grid points.
For instance, the non-linear term $a=bc$ (where $a,b,c$ are 3D arrays) can be computed as follows:
\begin{lstlisting}
!$acc kernels
do i=1,nx
 do j=1,nyp
  do k=1,nzp
   a(i,k,j)=b(i,k,j)*c(i,k,j)
  end do
 end do
end do
!$acc end kernels
\end{lstlisting}
where \texttt{nx, nyp} and \texttt{nzp} are the pencil dimensions in physical space.
For this part of the code, the use of the \texttt{kernels} construct or \texttt{parallel loop} construct exhibit very similar performance.
In addition, simple assignment operations (e.g. during time integration) are also performed using the \texttt{kernels} construct; this avoids expensive back-and-forth copies from CPU to GPU memories and vice-versa, which are limited by the available bandwidth between host and device.

Finally, the solution of the systems of Helmholtz problems is also offloaded to GPUs.
This step requires solving a series of tridiagonal systems using Gauss elimination along the wall-normal direction for each ($x,y$) position.
Thus, the only direction along which dependencies are present is the wall-normal direction while it can be parallelized along the $x$ and $y$ directions of the pencil.
Thanks to the adoption of the managed memory feature, the offloading to GPUs of this part of the code is rather straightforward and does not pose particular challenges or require extensive code modifications with respect to the CPU version.
 
\subsubsection{MPI communications}

\begin{figure}
\center 
\setlength{\unitlength}{0.0025\columnwidth}
\begin{picture}(400,140)
\put(50,0){\includegraphics[width=0.7\columnwidth, keepaspectratio]{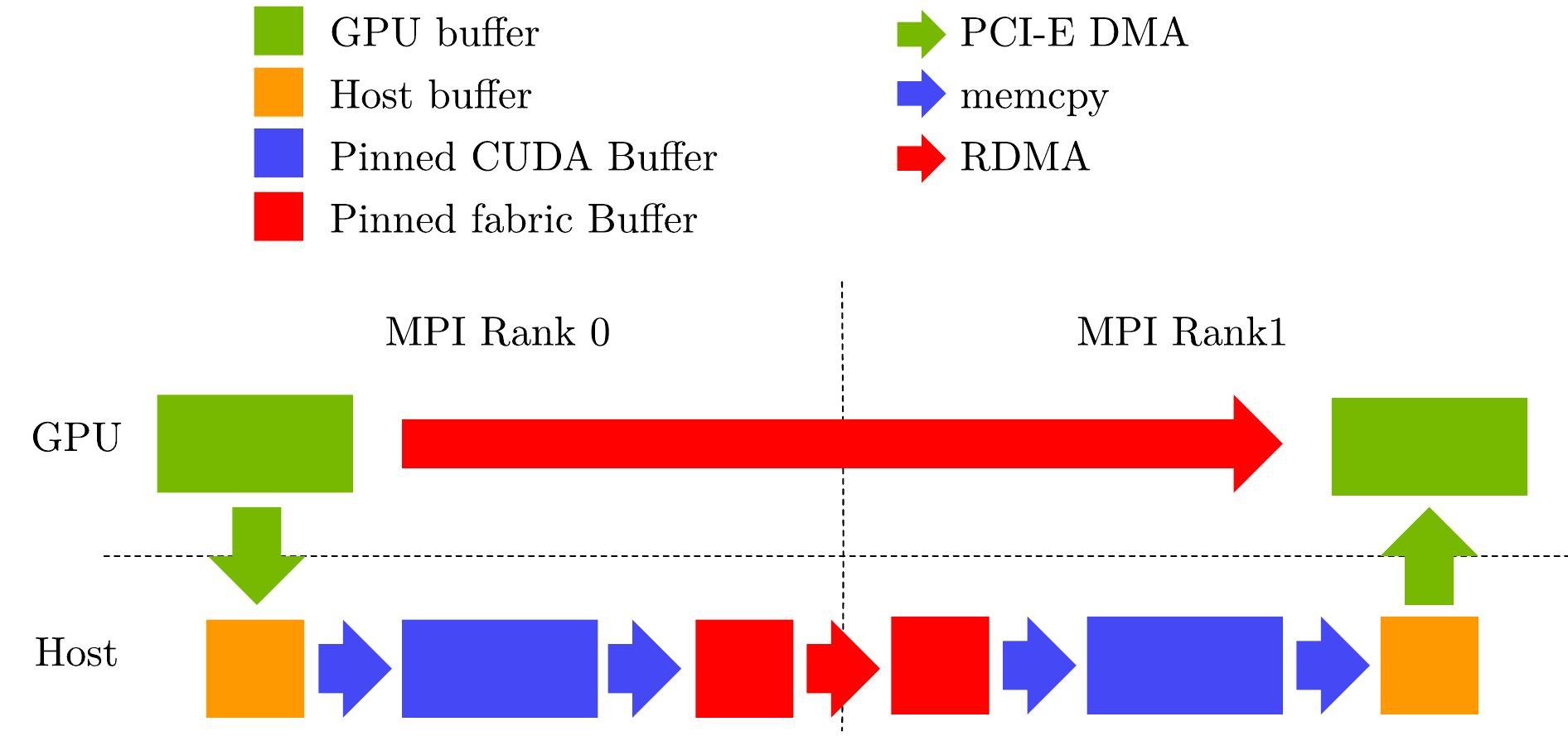}}
\end{picture}
\caption{Schematic showing the step required to perform an MPI communication between data stored in the buffer of GPU~0 and GPU~1.
In standard MPI installations, the data computed on GPU~0 has to be moved into host memory of CPU~0 and then sent to the MPI process 0 and, in turn, transferred to the GPU~1.
Using a CUDA-aware MPI implementation that exploit GPUDirect technologies, e.g. Remote Direct Memory Access (RDMA), a direct transfer between GPU~0 and GPU~0 can be performed (long red arrow). 
Reproduced from \cite{cudawmpi}.}
\label{aware}
\end{figure}

Once all the transforms and solver execution have been ported to the GPUs, as illustrated above, to improve code scalability, MPI communications can be also optimized.
Specifically, in traditional MPI installations (Figure~\ref{aware}), considering a message between rank 0 and rank 1, the data computed on GPU~0 has to be staged into the host memory of CPU~0 (green and blue arrows) and then sent to the MPI process 0 via MPI message and, in turn, to the GPU~1 (assigned to rank~1), as shown in Figure~\ref{aware}.
These additional passages represent a big overhead and can negatively impact strong and weak scalability results, especially when large-scale simulations are performed (i.e. using more than 100 GPUs) and many non-linear terms have to be computed (which require pencil transpositions and thus loop of MPI communications).
A possible solution to this problem is the use of CUDA-aware MPI implementations (or more in general GPU-aware).

To address this problem, we take advantage of the CUDA-aware MPI.
CUDA-aware MPI implementations allow for the use of GPUDirect, a group of technologies that provide high-bandwidth, low-latency communications for all kinds of inter-rank communication (intra-node, inter-node, and inter-node) via Remote Direct Memory Access (RDMA) when Nvidia GPUs are used.
In particular, when GPUDirect RDMA is available the GPU buffer can be directly moved to the network without passing through the host memory at all. 
So the data is transferred from the GPU buffer of the MPI Rank 0 to the GPU buffer of MPI Rank 1 (as represented by the long solid arrow in the upper part of the figure).
The use of CUDA-aware MPI features requires however the use of pinned buffers.
As here data movement is implicitly handled via the use of managed memory, to exploit CUDA-aware features, CUDA Fortran instructions are required to define the buffers used to perform the communications as pinned (thus inhibiting the memory paging performed by the managed memory feature).
This can be specified via the \texttt{pinned} attribute in the variable declarations and enabling the support for CUDA among the compiling options.
Then, via OpenACC directives, the use of CUDA-aware features can be specified as follows:
\begin{lstlisting}
!$acc data copy(bufs,bufr)
!$acc host_data use_device(bufs,bufr)
call mpi_isend(bufs,.....)
call mpi_irecv(bufr,......)
call mpi_waitall(.....)
!$acc end host_data
!$acc end data
\end{lstlisting}

where \texttt{bufr} and \texttt{bufs} are the two GPU pinned buffers used for the MPI communications.
It is convenient to allocate these buffers during code start-up as the allocation/deallocation of pinned buffers has a non-negligible overhead.
Also, it is worth mentioning that the use of non-blocking MPI communications is advisable as it allows the MPI library to build more efficient pipelines.
Finally, as the code relies on the managed memory feature (i.e. CUDA Unified memory), the chosen MPI implementation should support the CUDA unified memory, e.g. OpenMPI and MVAPICH-2 \cite{gabriel2004open,panda2013mvapich}.
Indeed, the MPI implementation needs to know where the data is located and which is the optimal strategy to perform the communication \cite{manian2019characterizing}.

\subsection{Optimization, performance and scaling}

Thanks to the use of OpenACC directives, the code has been ported to GPU using a step-by-step approach.
First, all the Fourier and Chebyshev transforms, which represent the most computationally intensive part of the algorithm and then the solver and the CUDA-aware MPI communications have been ported to GPUs.
During each of these steps, the code has been optimized and profiled using the tools provided by the NVIDIA Nsight Systems and using the NVIDIA Tools Extension (NVTX) instructions.
Specifically, NVTX instructions have been used to annotate the profiler timeline with events and ranges.
This procedure allows to find, and possibly remove, performance bottlenecks as well as to improve specific sections of the code.

To evaluate the performance of the GPU-porting, we first consider a single-node run where all the GPUs present in the node are used.
This allows us to obtain a first estimate of the gain that can be obtained by using GPUs, compared to using all the physical cores available on the CPU counterpart.
It is also possible to compare the results keeping fixed the number of MPI tasks used; however, this would underestimate CPU performance as only a part of the available cores will be used: the number of GPUs present in a computing node (from 4 to 8) is usually smaller than the number of cores available (from 16 up to 256).
For this test case, we consider a single-phase turbulent channel flow solved on a grid having $N_x \times N_y \times N_z = 256 \times 256 \times 257$ grid points. 
Tests have been performed on four different machines, one local server, two machines present at CINECA (Marconi-100 and Leonardo \cite{turisini2023leonardo}) and the CPU partition of LUMI (LUMI-C).
This latter machine is included as a reference of the performance that can be obtained in a CPU-based computing infrastructure.
The technical specifications of the four machines are reported in Table~\ref{spec}.
When CPUs are used, the maximum number of cores available is used whereas when GPUs are employed, the number of MPI tasks is set equal to the number of GPUs.
For all tests, the code is compiled with the Nvidia Fortran compiler \texttt{nvfortran} when available and when not with the GNU compiler \texttt{gfortran}.


\begin{table}[h]
\caption{Technical details of the computing infrastructures employed in the tests}
\begin{tabular*}{\hsize}{@{\extracolsep{\fill}}lll@{}}
\toprule
System & CPU & GPU\\
\colrule
Local Server & 1 x  Intel Xeon 5218 16 cores &  2 x Quadro RTX 5000 16 GB\\
Marconi-100 (CINECA)  & 2 x IBM POWER9 AC922 16 cores  &  4 x NVIDIA Volta V100 16 GB\\
Leonardo (CINECA) & 1 x  Intel Xeon 8358 32 cores &  4 x Nvidia Ampere A100 64GB\\
LUMI-C (LUMI) &  2 x AMD EPYC 7763 64 cores&  -\\

\botrule
\end{tabular*}
\label{spec}
\end{table}

\begin{figure}
\center 
\setlength{\unitlength}{0.0025\columnwidth}
\begin{picture}(400,155)
\put(50,0){\includegraphics[width=0.7\columnwidth, keepaspectratio]{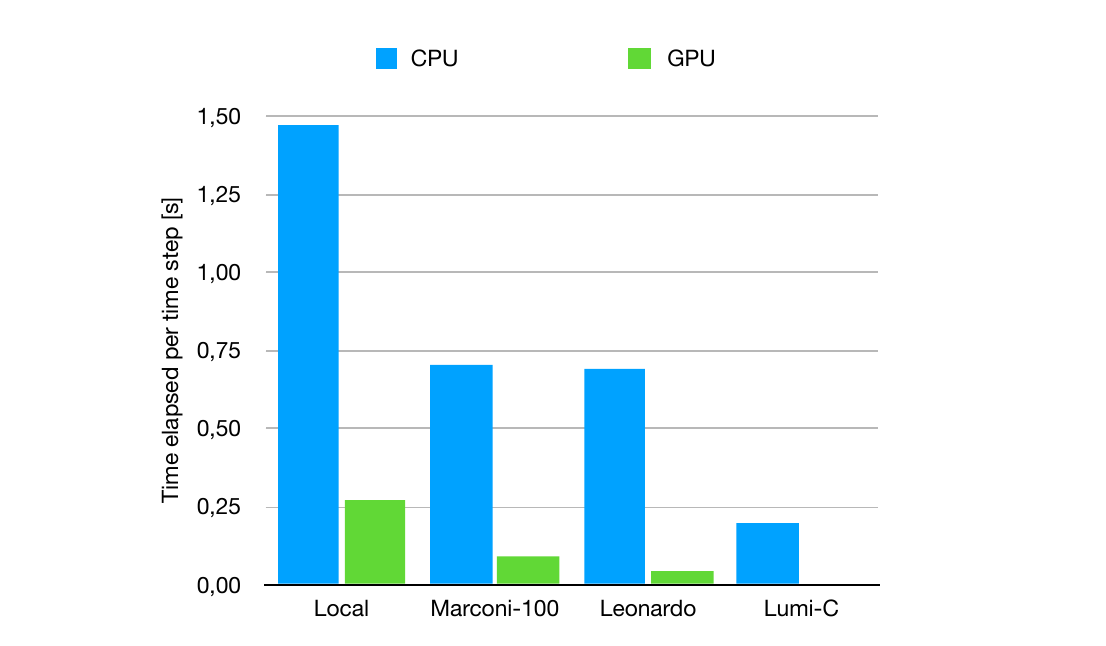}}
\end{picture}
\caption{Time elapsed per time step on different machines using all the physical cores available (blue) and all the GPUs available (green) on a single-node.
The results have been obtained considering a single-phase turbulent channel flow and a grid resolution equal to $N_x \times N_y \times N_z = 256 \times 256 \times 257$. 
For all cases, the code has been compiled using the Nvidia Fortran compiler \texttt{nvfortan} with or without the support for GPU-acceleration, depending on the case considered (CPU or GPU).
When the \texttt{nvfortan}  compiler is not available (LUMI-C), the code has been compiled using \texttt{gfortran}.}
\label{speed}
\end{figure}

The time elapsed for a single time step is shown in Figure~\ref{speed} for the different machines and cases considered.
Results obtained from the CPU runs are reported in blue while those obtained from GPU runs are in green.
In general, it can be observed that when the full node performance is compared, i.e. all physical cores versus all available GPUs, using the GPUs a consistent speed-up is obtained.
The specific value of the speed-up obtained depends on the machine considered: generally speaking, this value ranges from 4 (local server) up to 16 (Leonardo).
This variation is due to the different performance offered by the CPUs and GPUs present in the computing node.
Comparing the results with the CPU runs performed on LUMI-C, it is interesting to observe that the wall-clock time elapsed per time step obtained using GPUs is smaller.

\begin{figure}
\center 
\setlength{\unitlength}{0.0025\columnwidth}
\begin{picture}(400,160)
\put(25,0){\includegraphics[width=0.8\columnwidth, keepaspectratio]{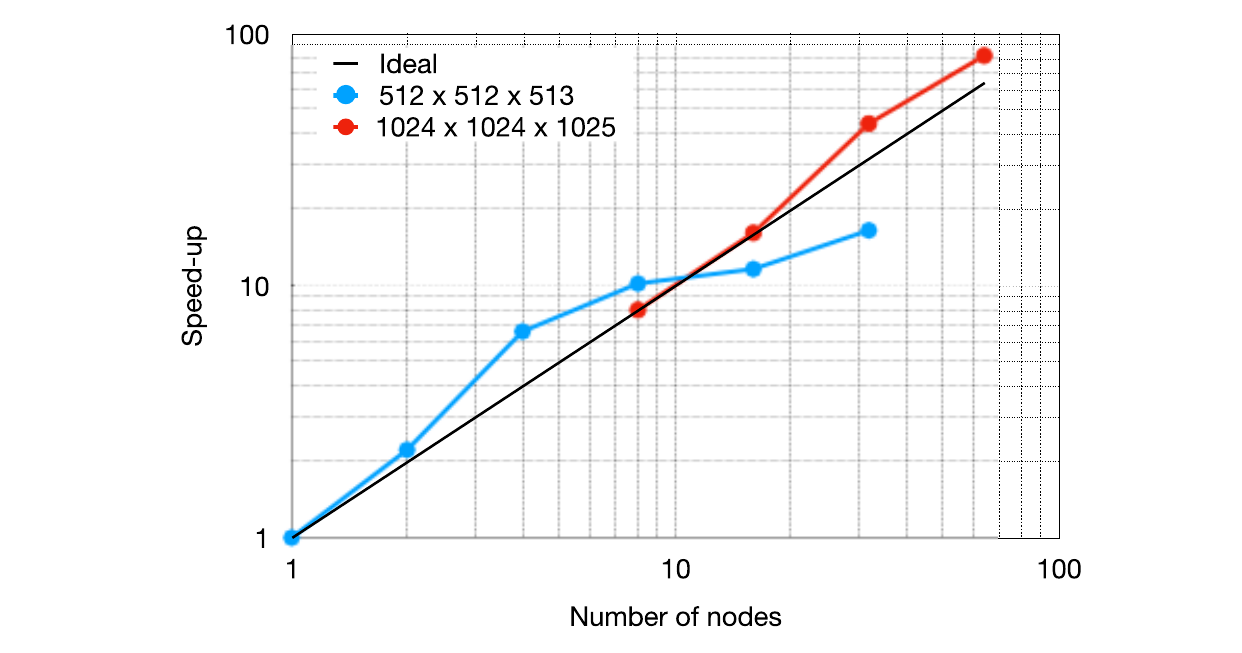}}
\end{picture}
\caption{Strong scaling results for the code FLOW36 obtained on Marconi100. 
Two different problems sizes have been considered: $N_x \times N_y \times N_z = 512 \times 512 \times 513$ and $N_x \times N_y \times N_z = 1024  \times 1024 \times 1025$.
For the smaller problem size (blue dots), tests have been performed starting from 1 node (4 GPUs) up to 32 nodes (128 GPUs) while, for the larger problem size (red dots), starting from 8 nodes (32 GPUs) up to 64 nodes (256 GPUs).}
\label{scaling}
\end{figure}

We now move to analyze the strong scaling results.
In particular, Figure~\ref{scaling} shows the results obtained employing Marconi-100, please refer to Table~\ref{spec} for the technical details of the cluster.
We consider two different grid resolutions, typical of production runs for direct numerical simulations of multiphase turbulence: $N_x \times N_y \times N_z = 512 \times 512 \times 513$ and $N_x \times N_y \times N_z = 1024  \times 1024 \times 1025$.
For the coarser grid resolution (blue dots), tests have been performed starting from 1 node (4 GPUs) up to 32 nodes (128 GPUs) while, for the finer grid resolution (red dots), starting from 8 nodes (32 GPUs) up to 64 nodes (256 GPUs).
A different number of nodes has been used for the two problem sizes because of the different memory requirements.
We can observe that in general good results are obtained for both the problem sizes considered.
For the smaller problem size, strong scaling results start to worsen when more than 16 nodes (64 GPUs) or more are used.
This is due to the lower computational load per pencil/slab and the relative cost of MPI communications.
Indeed, by increasing the problem size (i.e. considering the finer grid resolution), the computational load per pencil/slab increases and the cost of the MPI communications can be hidden; as a consequence, better results in terms of scaling are obtained.
Future developments are required to improve code scalability on a larger number of GPUs as performing large-scale 3D FFT require a significative amount of communications \cite{chatterjee2018scaling,czechowski2012communication,dalcin2019fast,pekurovsky2012p3dfft,ravikumar2019gpu}.

\section{Code capabilities}

\begin{figure}
\center 
\setlength{\unitlength}{0.0025\columnwidth}
\begin{picture}(400,170)
\put(30,0){\includegraphics[width=0.8\columnwidth, keepaspectratio]{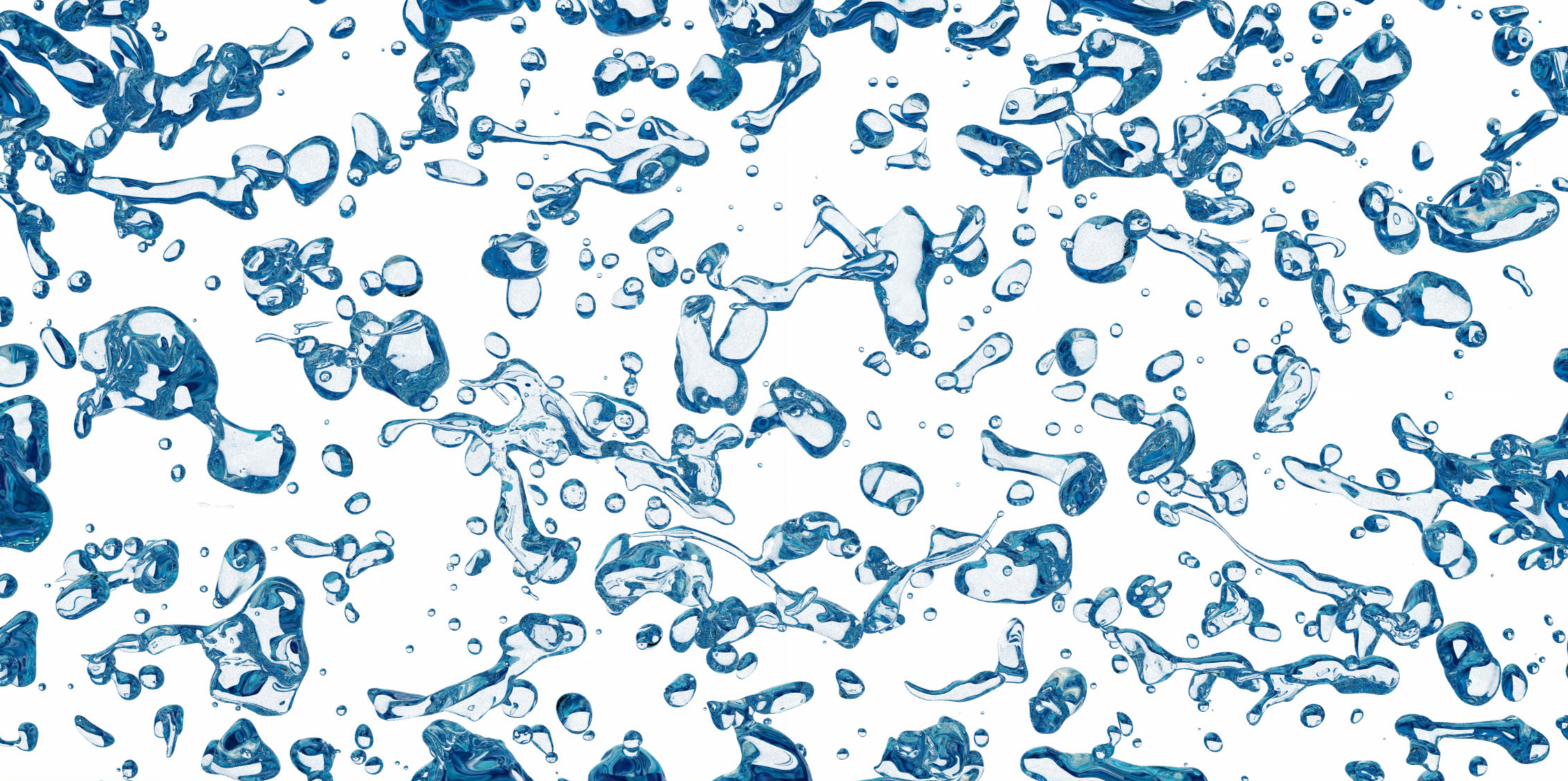}}
\end{picture}
\caption{Top view of a swarm of large and deformable drops released in a turbulent channel flow.
The flow moves from left to right (along the streamwise direction) and drops coalesce and break under the action of turbulence fluctuations.
The interface of the drops is identified as the iso-contour $\phi=0$.
The grid resolution employed for this demo simulation is $N_x=2048 \times 1024 \times 1025$.
This run has been executed using 64 nodes (256 GPUs) of Leonardo \cite{turisini2023leonardo}.}
\label{demo}
\end{figure}

We provide here an example of the code capabilities.
In particular, we consider the injection of a swarm of large and deformable drops released in a turbulent channel flow, a flow configuration similar to the one adopted in previous works \cite{Mangani2024,Roccon2017,Soligo2019c}.
The computational domain is a closed channel with dimensions $L_x \times L_y \times L_z = 4\pi h \times 2 \pi h \times  2h$ corresponding to $L_x^+ \times L_y^+ \times L_z^+ = 3770 \times 1885 \times 600$ wall units. Equations are discretized on a grid with $N_x \times N_y \times N_z = 2048 \times 1024 \times 1025$ collocation points. 
The simulation is performed at a fixed shear Reynolds number of $Re_\tau = 300$ and the Weber number has been set equal to $We=6$.
The simulation starts by releasing 256 spherical droplets in a flow field previously obtained from a single-phase flow simulation of a fully developed turbulent channel flow.
After an initial transient, where drops start to break and coalesce following complex dynamics, a new equilibrium situation is reached in which a balance between coalescence and breakage events is attained.
Figure~\ref{demo} shows a qualitative rendering of the steady-state configuration attained by the system at $t^+=3000$.
The flow moves from left to right (along the streamwise direction $x$) and a top view of the system is shown (streamwise-horizontal; spanwise-vertical).
The interface of the drops is identified as the iso-contour $\phi=0$ and is rendered using a ray tracing algorithm.
We can observe the wide range of scales and shapes that characterize the drops: from very small and almost undeformed spherical droplets to very large drops characterized by a complex three-dimensional shape.
It is also worth noticing the presence of elongated filaments/threads that will eventually lead to the formation of small drops.

\section{Conclusions and future developments}

We detailed the GPU-porting of a pseudo-spectral code tailored towards large-scale simulations of multiphase flow, more specifically interface-resolved direct numerical simulations.
The code relies on direct numerical simulation of turbulence coupled with a phase-field method to describe the interface shape, deformation, and topological changes.
The development of the code has been designed with the specific goal of simplifying code maintenance, obtaining a portable code that can exploit GPU acceleration, and facilitating the implementation of new modules and new physics (e.g. additional governing equations). 
To achieve this ambitious goal, we rely on two levels of parallelism: i) A first level that relies on MPI and a 2D domain decomposition to divide the workload among the MPI tasks,; ii) A second level that relies on OpenACC directives and CUDA libraries (cuFFT) to accelerate code execution on GPU-based computing infrastructures.
The code makes use of the managed feature (CUDA Unified memory) to simplify code maintenance and avoid the explicit definition of data transfers between CPU and GPU memories \cite{lindholm2008nvidia,negrut2014unified}
In addition, pinned memory buffers are used to perform MPI communications so that the GPU Direct technologies can be employed to improve latency and data transfer times.
The main limitation of the employed approach is the compatibility with GPUs from other vendors as the managed memory feature is currently available only on the Nvidia HPC-SDK environment.
However, a similar feature has been recently released in the context of AMD GPU architectures \cite{chu2013amd,jin2022evaluating}.
Hence, in future developments, we plan to extend the support also to AMD GPUs by exploiting the unified memory feature present in the AMD ecosystem and combining it with openMP directives and rocFFT libraries to accelerate the code execution (solver and transforms) \cite{rocFFT}.

\section*{Acknowledgements}

The author would like to thank Giovanni Soligo, who developed the CPU-version of the present code.
We acknowledge the CINECA award under the ISCRA initiative, for the availability of high-performance computing resources and support (project HP10CH4PTZ and HP10BUJEO5).
We also acknowledge the EuroHPC Joint Undertaking for awarding this project access (PRACE-DEV-2022D01-069) to the EuroHPC supercomputer Marconi-100 and Leonardo, hosted by CINECA (Italy).
The authors gratefully acknowledge financial support from the European Union–NextGenerationEU.





\bibliography{totalbib}
\bibliographystyle{elsarticle-harv}


\end{document}